%% file: charm2015_APalanoBabar.tex
\documentclass[12pt]{article}
\usepackage{graphicx}
\usepackage{xspace}
\usepackage[percent]{overpic}
\usepackage{colordvi}
\usepackage{xcolor}
\newcommand\pubnumber{}
\newcommand\pubdate{\today}

\def\palano{INFN and University of Bari, Italy}
\def\support{\footnote{on behalf of the \babar\ Collaboration}}
\def\supporta{\footnote{Work supported by INFN, Italy and Thomas Jefferson National Accelerator Facility, 12000 Jefferson Avenue, Newport News, VA 23606, USA}}
\def\Title#1{\begin{center} {\Large #1 } \end{center}}
\def\Author#1{\begin{center}{ \sc #1} \end{center}}
\def\Address#1{\begin{center}{ \it #1} \end{center}}

\newcommand\pubblock{\rightline{\begin{tabular}{l} \pubnumber\\
         \pubdate  \end{tabular}}}
\newenvironment{Abstract}{\begin{quotation}  }{\end{quotation}}
\newenvironment{Presented}{\begin{quotation} \begin{center} 
             PRESENTED AT\end{center}\bigskip 
      \begin{center}\begin{large}}{\end{large}\end{center} \end{quotation}}

\input econfmacros.tex
\def\epem       {\ensuremath{e^+e^-}\xspace}
\def\Kp    {\ensuremath{K^+}\xspace}
\def\Km    {\ensuremath{K^-}\xspace}

\def\KpKm  {\ensuremath{\Kp \kern -0.16em \Km}\xspace}
\def\KS    {\ensuremath{K^0_{\scriptscriptstyle S}}\xspace} 
\def\piz   {\ensuremath{\pi^0}\xspace}

\def\pip   {\ensuremath{\pi^+}\xspace}
\def\pim   {\ensuremath{\pi^-}\xspace}

\def\etac     {\ensuremath{\eta_c}\xspace}

\def\calR         {{\ensuremath{\cal R}\xspace}}

\def\jpsi  {\ensuremath{J/\psi}\xspace}

\newcommand{\kkpiz}{\ensuremath{\Kp\Km \piz}\xspace}
\newcommand{\gevc}{\ensuremath{{\mathrm{\,Ge\kern -0.1em V\!/}c}}\xspace}
\newcommand{\mevc}{\ensuremath{{\mathrm{\,Me\kern -0.1em V\!/}c}}\xspace}
\newcommand{\gevcc}{\ensuremath{{\mathrm{\,Ge\kern -0.1em V\!/}c^2}}\xspace}
\newcommand{\mevcc}{\ensuremath{{\mathrm{\,Me\kern -0.1em V\!/}c^2}}\xspace}
\def\BR         {{\ensuremath{\cal B}\xspace}}

\newcommand{\mev}{\ensuremath{\mathrm{\,Me\kern -0.1em V}}\xspace}

\newcommand{\MM}{\ensuremath{M^2_{\mathrm{rec}}}\xspace}

\newcommand{\pipiz}{\ensuremath{\pip \pim \piz}\xspace}
\newcommand{\psipipiz}{\ensuremath{\jpsi \ \to \ \pip\pim \piz}\xspace}
\newcommand{\psikkpiz}{\ensuremath{\jpsi \ \to \ \Kp\Km \piz}\xspace}

\def\calR         {{\ensuremath{\cal R}\xspace}}
\usepackage{relsize}
\def\babar{\mbox{\slshape B\kern-0.1em{\smaller A}\kern-0.1em
    B\kern-0.1em{\smaller A\kern-0.2em R}}}
\begin{document}
\begin{titlepage}
\pubblock

\vfill
\Title{Dalitz plot analysis of three-body Charmonium Decays at \babar}
\vfill
\Author{Antimo Palano\support\supporta}
% put in address(es) defined above
\Address{\palano}
\vfill
\begin{Abstract}
We present preliminary results on the measurement of the I=1/2 $K \pi$ $\mathcal{S}$-wave through a model independent partial wave analysis of
$\eta_c$ decays to $\KS \Kp \pim$ and $\Kp \Km \piz$ produced in two-photon interactions. We also perform a Dalitz plot analysis
of the $J/\psi$ decays to $\pip \pim \piz$ and $\Kp \Km \piz$ produced by the initial state radiation process.
\end{Abstract}
\vfill
\begin{Presented}
The 7th International Workshop on Charm Physics (CHARM 2015)\\
Detroit, MI, 18-22 May, 2015
\end{Presented}
\vfill
\end{titlepage}
\def\thefootnote{\fnsymbol{footnote}}
\setcounter{footnote}{0}
%
%%%%%%%%%%%%%%%%%%%%%%%%%%%%%%%%%%
\section{Introduction}
%$\Box$}
Charmonium decays can be used to obtain new information on light meson spectroscopy.
  In $e^+ e^-$ interactions, samples of charmonium decays can be obtained using different processes.
  \begin{itemize}
  \item{}
  In two-photon interactions we select events in which
 the $e^+$ and $e^-$  beam particles are scattered at small
 angles and remain undetected.
  Only resonances with $J^{PC}=0^{\pm+}, 2^{\pm+}, 3^{++}, 4^{\pm+}$....
 can be produced.
\item{}
 In the Initial State Radiation (ISR) process,
we reconstruct events having a (mostly
undetected) fast forward $\gamma_{ISR}$ and only $J^{PC}=1^{--}$ states can be produced.
  \end{itemize}

\section{Study of $\eta_c \to K \bar K \pi$}
The BaBar Dalitz plot analysis of the $\eta_c \to K^+ K^- \eta$ and $\eta_c \to K^+ K^- \piz$ has provided the unexpected observation of $K^*_0(1430)\to K \eta$~\cite{babar1}.
 We also find that the $\eta_c$ three-body hadronic decays proceed almost entirely through the
 intermediate production of scalar meson resonances.

 We study the reactions~\cite{conj}
\begin{center}
  $\gamma \gamma \to  \KS \Kp \pim$,\\ 
  $\gamma \gamma \to  \Kp \Km \piz $ 
\end{center}

\begin{figure}
  \centering
    \begin{overpic}[clip,height=2.5in]{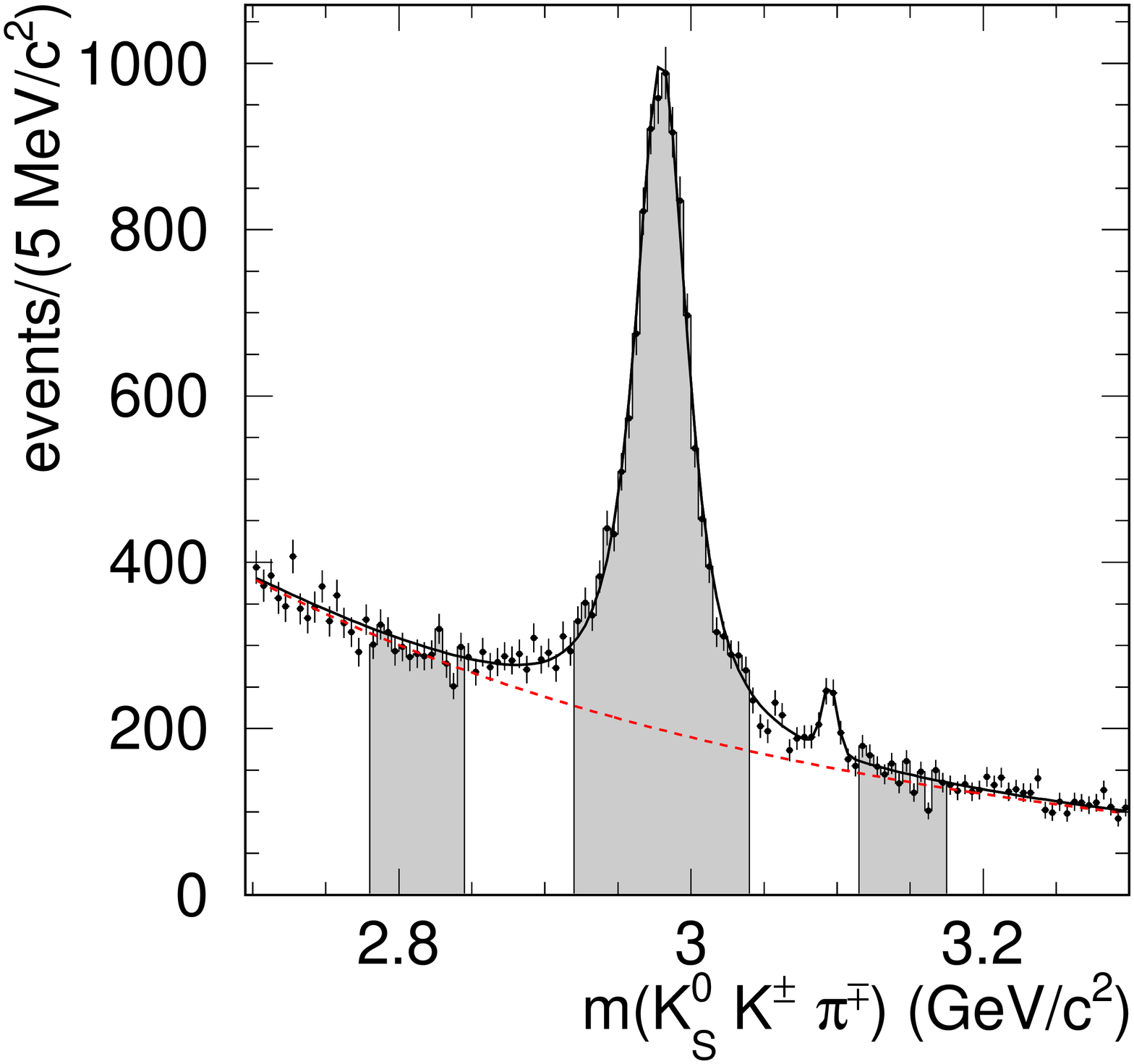}
      \put(65,75){\scriptsize\color{black}\babar}
      \put(58,70){\tiny\color{black}preliminary}
    \end{overpic}
    \begin{overpic}[clip,height=2.5in]{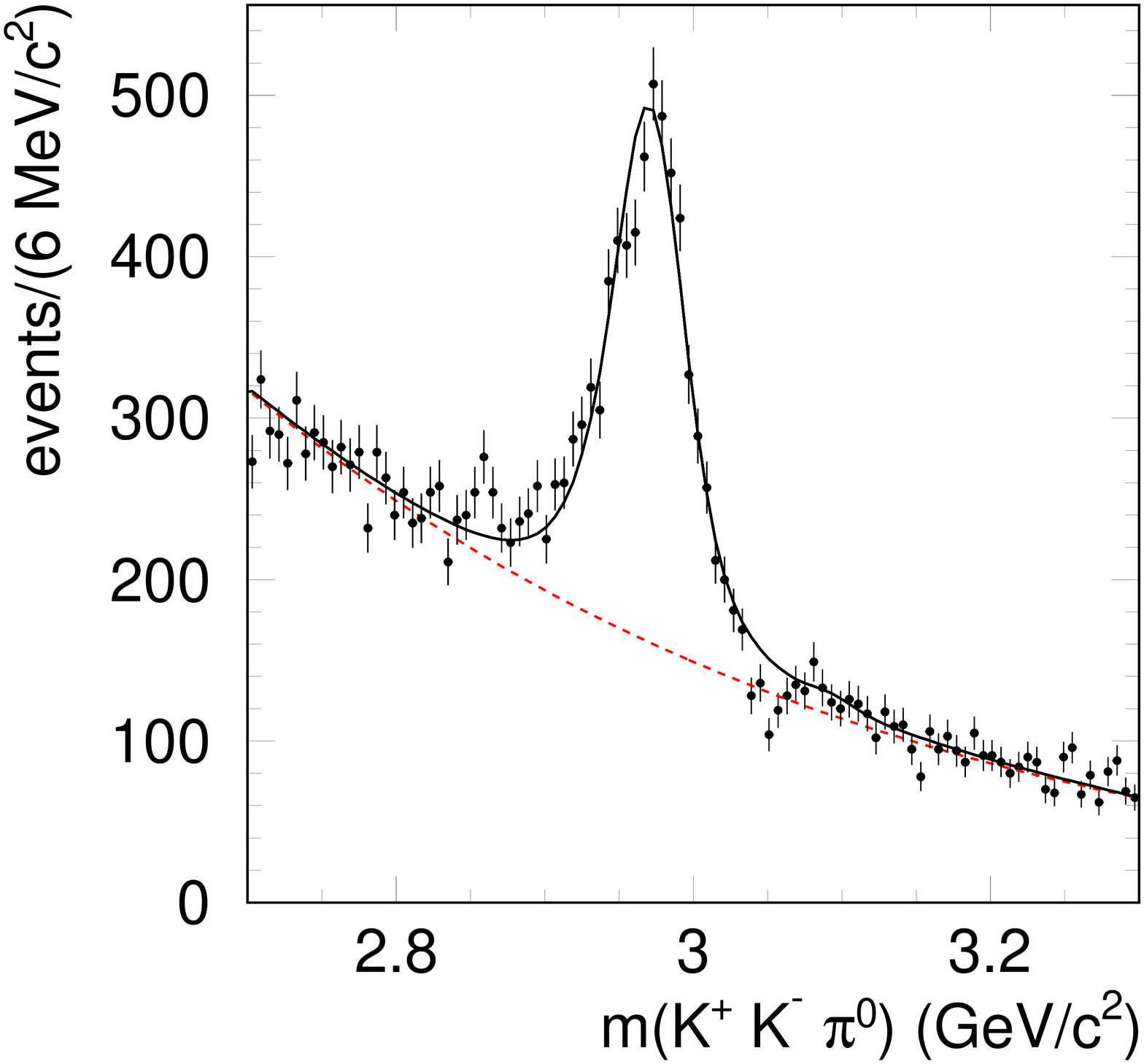}
\end{overpic}
\caption{(Left) $\KS \Kp \pim$ and (Right) $\Kp \Km \piz$ mass spectra in two-photon interactions.}
\label{fig:fig1}
\end{figure}
In the following, details on events reconstruction will be given only for the $\KS \Kp \pim$ final state.
We select events having only four tracks. Since two-photon events balance the transverse momentum, we require $p_T$, the transverse momentum of the $\KS \Kp \pim$ system with respect to the beam axis, to be $p_T<0.08 \ GeV/c$.
We also define $\MM\equiv(p_{\epem}-p_{\mathrm{rec}})^2$, where $p_{\epem}$ is the four-momentum of the initial state and $p_{\mathrm{rec}}$ is the four-momentum of the $\KS \Kp \pim$ system and remove ISR events requiring  $\MM>10\ GeV^2/c^4$.
The $K \bar K \pi$ mass spectra in the $\eta_c$ mass regions are shown in Fig.~\ref{fig:fig1}.

\begin{figure}
\centering
        \begin{overpic}[clip,height=2.5in]{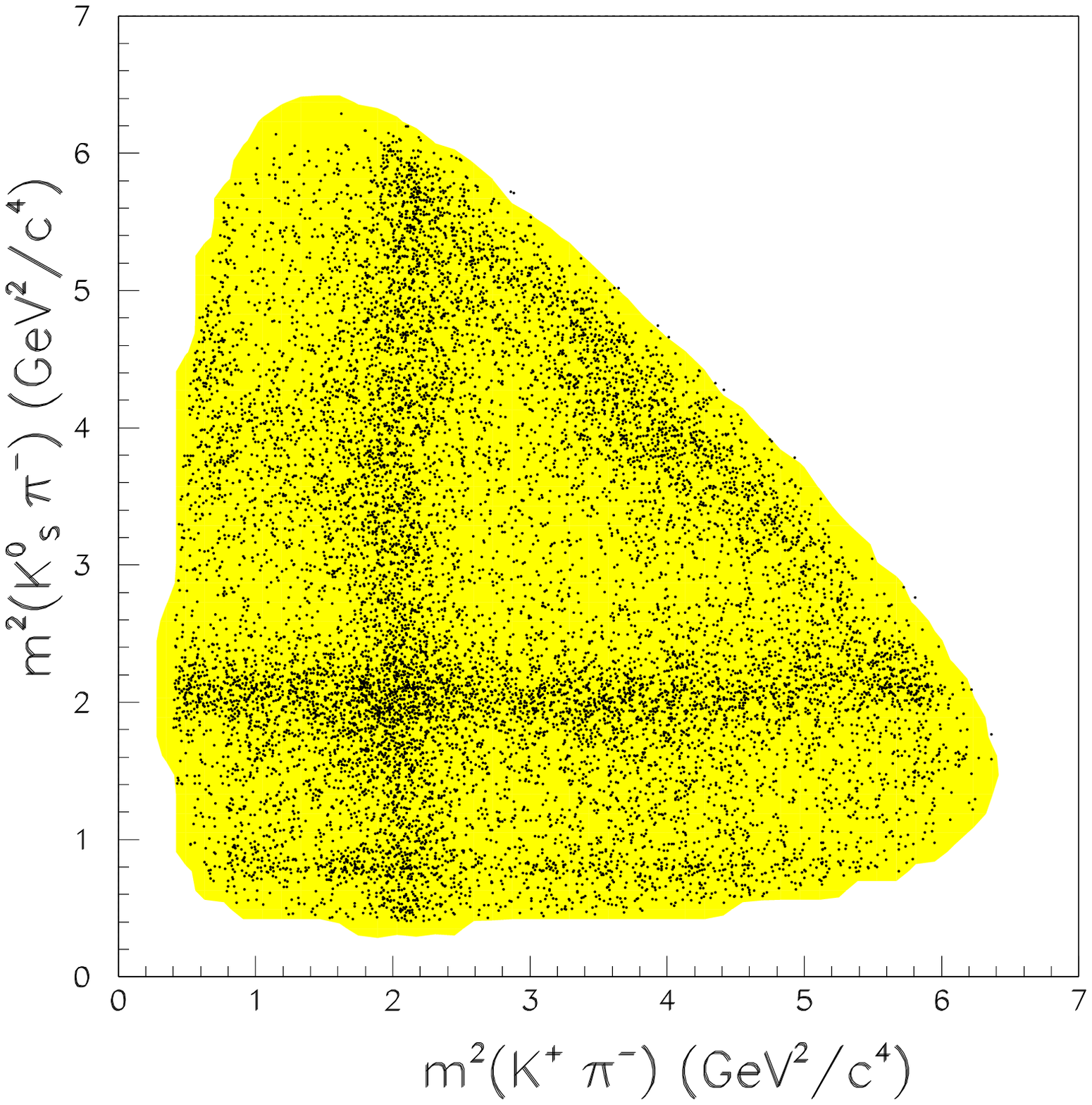}
          \put(65,80){\scriptsize\color{black}\babar}
          \put(58,75){\tiny\color{black}preliminary}
    \end{overpic}          
    \includegraphics[height=2.5in]{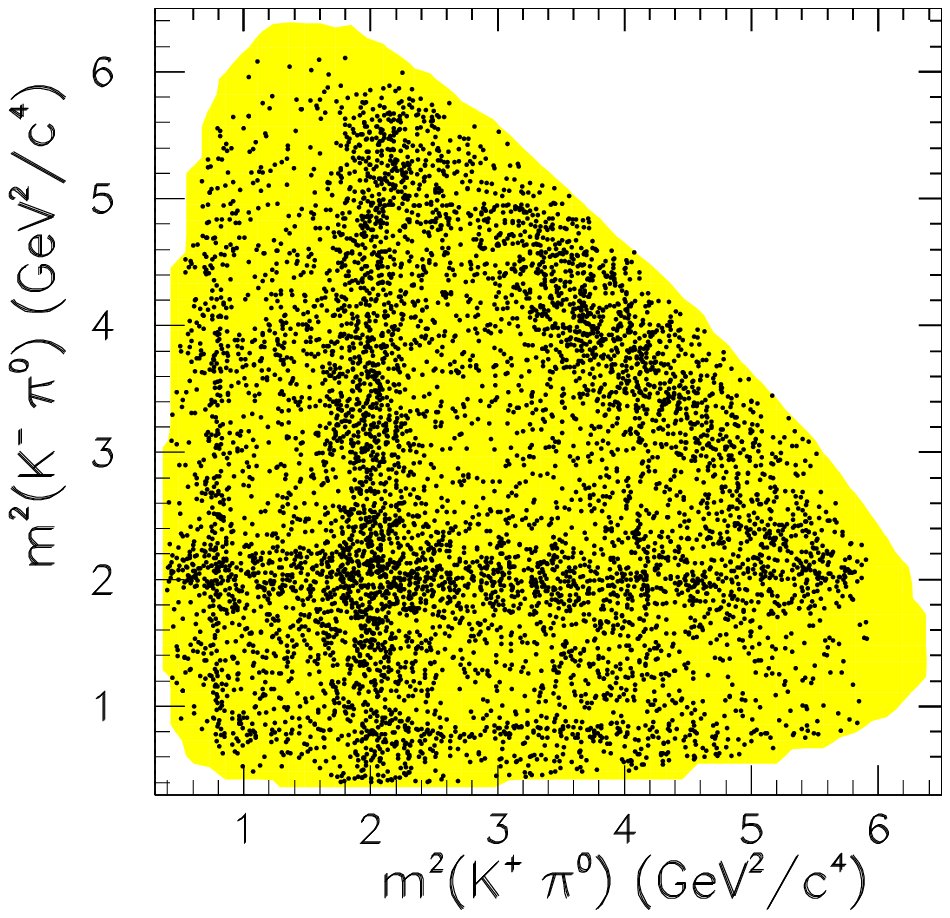} 
\caption{(Left) $ \eta_c \to \KS \Kp \pim$ and (Right) $ \eta_c \to \Kp \Km \piz$ Dalitz plots.}
\label{fig:fig2}
\end{figure}

The $\eta_c$ signal regions contain 12849 events with (64.3 $\pm$ 0.4)\% purity for  $ \eta_c \to \KS \Kp \pim$ and
6494 events with (55.2$\pm$0.6)\% purity for $ \eta_c \to  \Kp \Km \piz$. 
The Dalitz plots for the two $\eta_c$ decay modes are shown in Fig.~\ref{fig:fig2} and are dominated by the presence
of $K^*_0(1430)$.

The backgrounds below the $\eta_c$ signals are estimated from the sidebands. We observe asymmetric $K^*$'s in the
background to the $ \eta_c \to \KS \Kp \pim$ final state due to interference between I=1 and I=0 contributions.

\section{Dalitz plot analysis of $\eta_c \to K \bar K \pi$}

We perform unbinned maximum likelihood fits using the Isobar model~\cite{asner} and Model Independent Partial Wave Analysis (MIPWA)~\cite{mipwa}.
In the MIPWA the $K \pi$ mass spectrum is divided into 30 equally spaced mass intervals 60 MeV wide and for each bin we add to the fit two new free parameters, the amplitude and the phase of the $K \pi$ $\mathcal{S}$-wave (constant inside the bin).
     We also fix the amplitude to 1.0 and its phase to $\pi/2$ in an arbitrary interval of the mass spectrum (bin 11 which corresponds to a mass of 1.42 \gevcc). The number of additional free parameters is therefore 58.
Due to isospin conservation in the decays, amplitudes are symmetrized with respect to the two $K \pi$ decay modes.
 The $K^*_2(1420)$, $a_0(980)$, $a_0(1400)$, $a_2(1310)$, ... contributions are modeled as relativistic Breit-Wigner functions multiplied by the corresponding angular functions.
 Backgrounds are fitted separately and interpolated into the $\eta_c$ signal regions.
 The fits improves when an additional high mass $a_0(1950) \to K \bar K$ I=1 resonance is included with free parameters in both $\eta_c$ decay modes. The weighted average of the two measurement is:
 $m(a_0(1950))=1931 \pm 14 \pm 22$ \mevcc, $\Gamma(a_0(1950))=271 \pm 22 \pm 29$ \mev. The statistical significances for the $a_0(1950)$ effect (including systematics) are $2.5\sigma$ for $\etac \to \KS \Kp \pim$ and $4.0\sigma$  for $\etac \to \Kp \Km \piz$.

\begin{figure}
\begin{center}
  \includegraphics[height=2.0in]{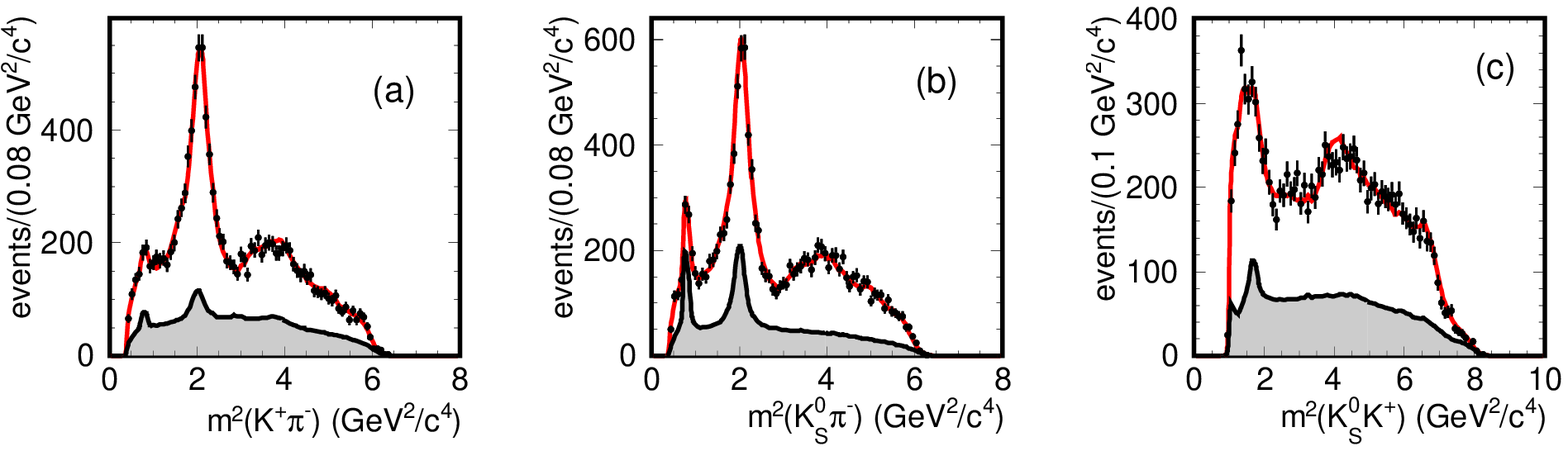}
  \includegraphics[height=2.0in]{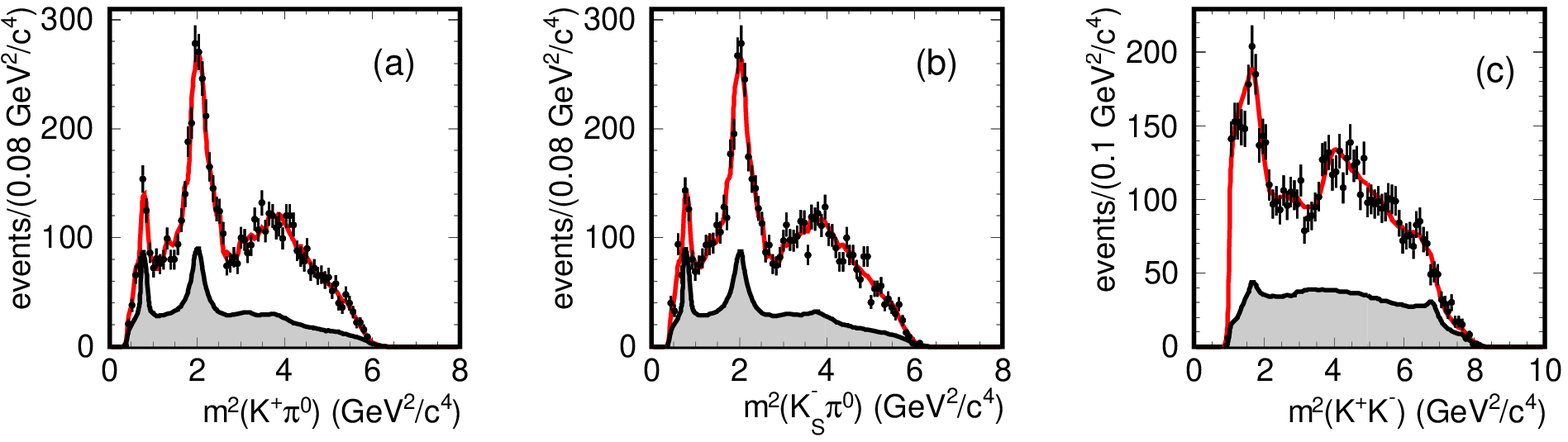}
\end{center}
\caption{(Top) $ \eta_c \to \KS \Kp \pim$ and (Bottom) $ \eta_c \to \Kp \Km \piz$ Dalitz plots projections. The superimposed curves are from the fit results. Shaded is contribution from the interpolated  background.}
\label{fig:fig3}
  \end{figure}

\begin{figure}
    \centering
      \begin{overpic}[clip,height=2.5in]{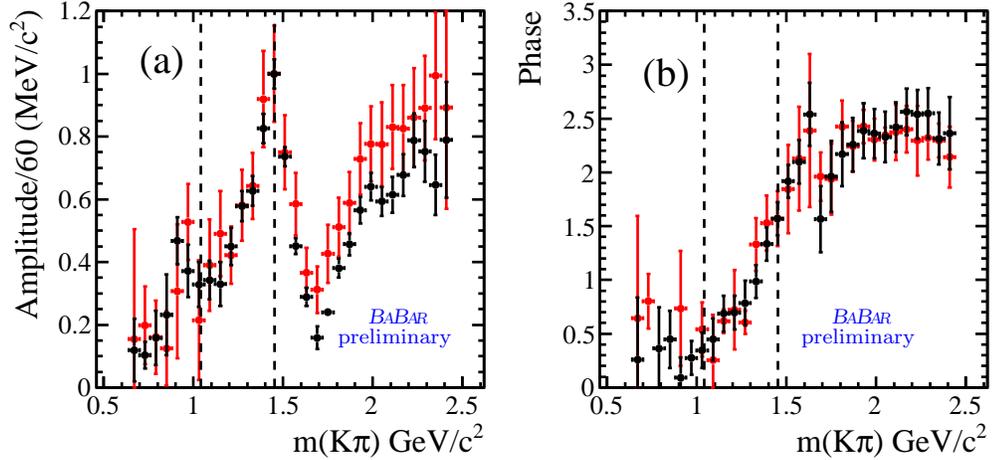}
      \put(36,14){\scriptsize\color{blue}\babar}
      \put(33,12){\scriptsize\color{blue}preliminary}
      \put(82,14){\scriptsize\color{blue}\babar}
      \put(79,12){\scriptsize\color{blue}preliminary}      
      \end{overpic}
     \caption{Fitted $K \pi$ $\mathcal{S}$-wave amplitude and phase from MIPWA. Red crosses are for $\eta_c \to \Kp \Km \piz$, black crosses are $\eta_c \to \KS \Kp \pim$. The dashed lines indicate the $K \eta$ and $K \eta'$ thresholds.}
\label{fig:fig4} 
  \end{figure}

The Dalitz plot projections with fit results for $\eta_c \to \KS \Kp \pim$ and $\etac \to \Kp \Km \piz$ are shown in Fig.~\ref{fig:fig3}. The fitted fractions and phases are given in Table~\ref{tab:tab2}. We observe e vary good fit to the data.
We note that the $K^*(892)$ contributions arise entirely from background.
    \begin{table}
\begin{center}
{\scriptsize
\begin{tabular}{lcccc}
  \hline
  &  \multicolumn{2}{c}{ $\etac \to \KS \Kp \pim$} & \multicolumn{2}{c}{ $ \etac \to \Kp \Km \piz$} \cr
Amplitude & Fraction (\%) & Phase & Fraction (\%) & Phase\cr
\hline
 $(K\pi \ \mathcal{S}$-wave)K &  107.3 $\pm$ 2.6 $\pm$ 17.9&  0. &  125.5 $\pm$ 2.4 $\pm$ 4.2 &  0.\cr
$a_0(980) \pi$ & 0.83 $\pm$ 0.46 $\pm$ 0.80& 1.08 $\pm$ 0.18  $\pm$ 0.18 & 0.00 $\pm$ 0.03 $\pm$ 1.7 & ..... \cr
$a_0(1450) \pi$ & 0.7 $\pm$ 0.2 $\pm$ 1.4 & 2.63 $\pm$ 0.13 $\pm$ 0.17 & 1.2 $\pm$ 0.4 $\pm$ 0.7 & 2.90 $\pm$ 0.12 $\pm$ 0.25\cr
$a_0(1950) \pi$ & 3.1 $\pm$ 0.4 $\pm$ 1.2 & $-$1.04 $\pm$ 0.08 $\pm$ 0.77& 4.4 $\pm$ 0.8 $\pm$ 0.7& $-$1.45 $\pm$ 0.08 $\pm$ 0.27\cr
$a_2(1320) \pi$& 0.15 $\pm$ 0.06 $\pm$ 0.08 & 1.85 $\pm$ 0.20 $\pm$ 0.23 & 0.61 $\pm$ 0.23 $\pm$ 0.3& 1.75 $\pm$ 0.23 $\pm$ 0.42\cr
$K^*_2(1430)^0 K$ & 4.7 $\pm$ 0.9 $\pm$ 1.4 & 4.92 $\pm$ 0.05 $\pm$ 0.1 & 3.0 $\pm$ 0.8 $\pm$ 4.4 & 5.07 $\pm$ 0.09 $\pm$ 0.3\cr
\hline
Total & 116.8 $\pm$ 2.8 & & 134.8 $\pm$ 2.7 & \cr
$\chi_2/N_{cells}$ & 301/254=1.17 & & 283.2/233=1.22 & \cr
\hline
\end{tabular}
}
\caption{Amplitudes and phases from the MIPWA fits to the two $\eta_c$ decay modes.}  
\label{tab:tab2}
\end{center}
    \end{table}
    In comparison, the isobar model gives a worse description of the data, with $\chi^2/N_{cells}$ of 457/254=1.82 and 383/233=1.63 respectively fro the two $\eta_c$ decay modes.
The resulting $K \pi$ $\mathcal{S}$-wave amplitude and phase for the two $\eta_c$ decay modes is shown in Fig.~\ref{fig:fig4}.
We observe a clear $K^*_0(1430)$ resonance signal with the corresponding expected phase motion.
At high mass we observe the presence of the broad $K^*_0(1950)$ contribution with good agreement between the two $\eta_c$ decay modes.
  Comparing with LASS~\cite{lass} and E791~\cite{mipwa} experiments we note that phases before the $K \eta'$
threshold are similar, as expected from Watson~\cite{watson} theorem but amplitudes are very different.

\section{ Dalitz plot analysis of \psipipiz\ and \psikkpiz}
  
     Only a preliminary result exists, to date, on a Dalitz-plot analysis of $J/\psi$ decays to $\pip \pim \piz$~\cite{bill}. While large samples of $J/\psi$ decays exist, some branching fractions remain poorly measured. 
 BES III experiment has performed an angular analysis of  $J/\psi \ \to \ \Kp \Km \piz$.
 The analysis requires the presence of a broad $J^{PC}=1^{--}$ state in the $K^+K^-$ threshold region, which is interpreted as a multiquark state~\cite{bes}.
 
\begin{figure}
   \centering
               \begin{overpic}[clip,height=2.5in]{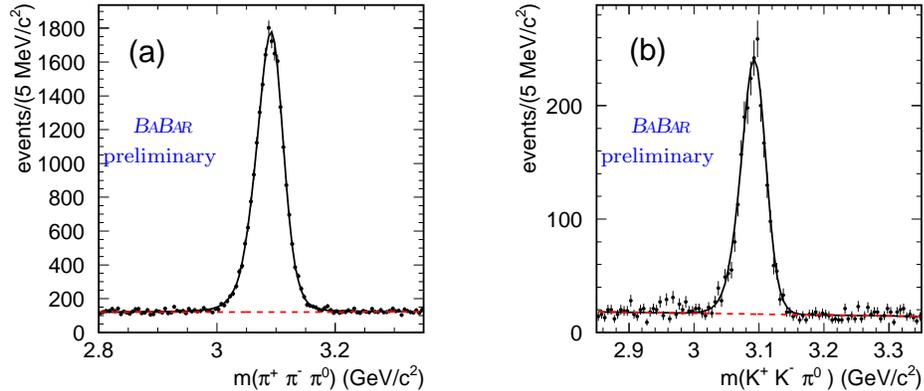}
      \put(15,30){\scriptsize\color{blue}\babar}
      \put(12,27){\scriptsize\color{blue}preliminary}
      \put(65,30){\scriptsize\color{blue}\babar}
      \put(62,27){\scriptsize\color{blue}preliminary}      
      \end{overpic}
      \caption{(Left) $\pip \pim \piz$ and (Right) $ \Kp \Km \piz$ mass spectra for ISR events candidates.}
\label{fig:fig5}
 \end{figure}

  We study the following reactions:
\begin{center}
$\epem \ \to \ \gamma_{\rm ISR} \ \pipiz, \ \qquad \epem \ \to \ \gamma_{\rm ISR} \ \kkpiz$
\end{center}
where $\gamma_{\rm ISR}$ indicate the (undetected) ISR photon.
 We select events having only two tracks and one (mass constrained) \piz.
 We compute $\MM \equiv ( p_{e^-}+p_{e^+}-p_{h^+}-p_{h^-}-p_{\pi^0})^2$, where $h=\pi/K$.
 This quantity should peak near zero for ISR events.
     We select events in the ISR region by requiring $|\MM|<2\ GeV^2/c^4$ and obtain the  \jpsi signals shown in Fig.~\ref{fig:fig5}.
 We fit the mass spectra using the Monte Carlo resolution functions described by a Crystal Ball+Gaussian functions. We obtain 21974 events for $\jpsi \ \to \ \pip \pim \piz$ with (86.1 $\pm$ 1.3)\% purity and 2393 for $\jpsi \ \to \ \Kp \Km \piz$
 with (87.8 $\pm$ 0.7)\% purity.
 The efficiency is mapped and fitted on the ($m(h^+ h^-), cos \theta_h$) plane, where $\theta_h$ is the $h^+$ helicity angle in the \jpsi rest frame. We weight each event by the inverse of the efficiency and perform background subtraction by
 assigning negative weights to events the $J/\psi$ sidebands regions.

 We obtain the following preliminary result:
  \begin{center}
$\calR = \frac{\BR(\jpsi \ \to \ \Kp \Km \piz)}{\BR(\jpsi \ \to \ \pip \pim \piz)} = 0.0929 \pm 0.002 \pm 0.002$
\end{center}
 The PDG reports $\BR(\jpsi \ \to \ \Kp \Km \piz)=55.2 \pm 0.12\times 10^{-4}$, based on 25 events, and 
  $\BR(\jpsi \ \to \ \pip \pim \piz)=2.11 \pm 0.07\times 10^{-2}$.
 These values give a ratio $\calR = 0.262 \pm 0.057$, which differs from our result by 3$\sigma$.
 
 \section{$\jpsi \ \to \ \pip \pim \piz$ Dalitz plot analysis}
\begin{figure}
   \centering
      \begin{overpic}[clip,height=2.5in]{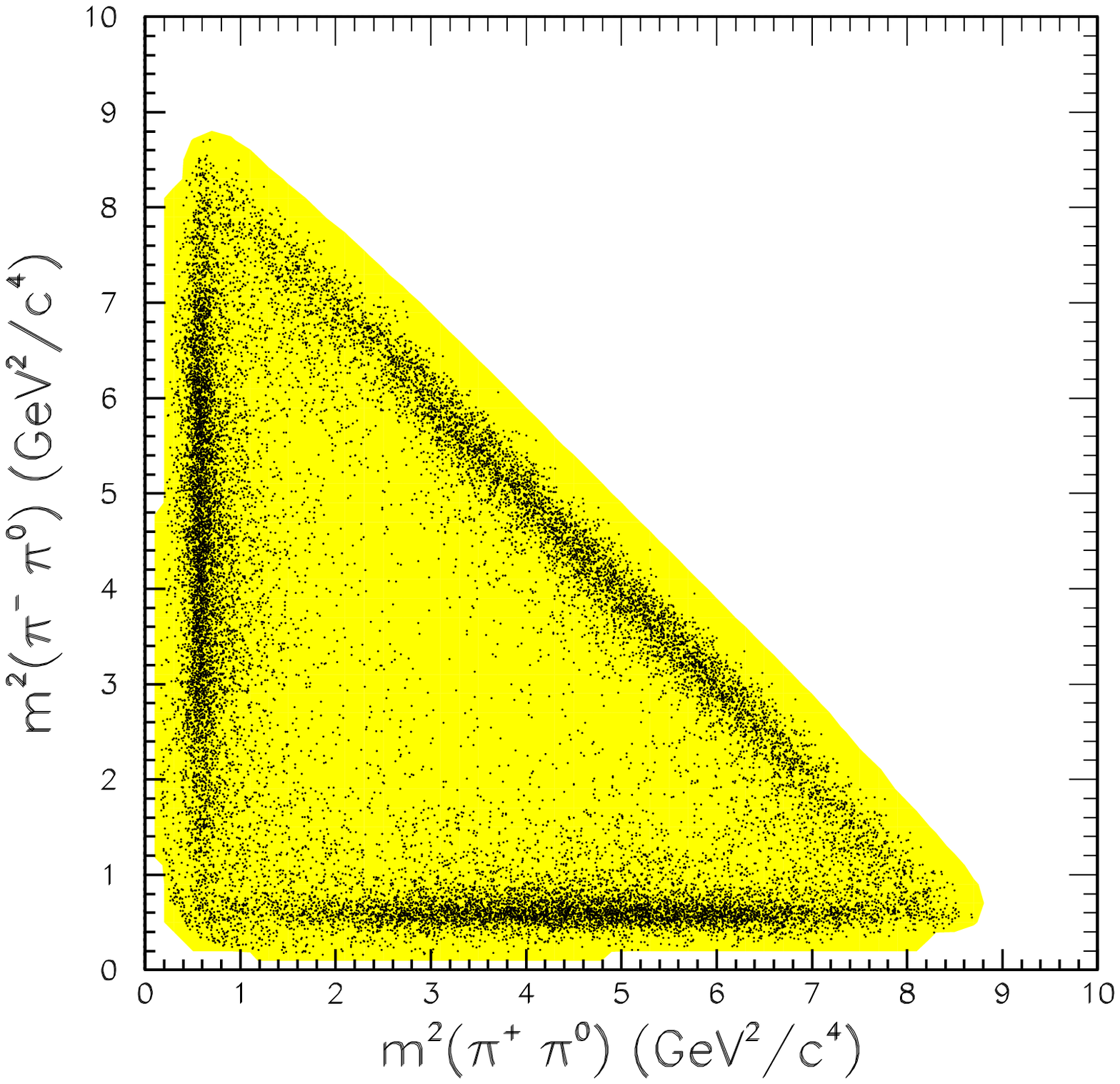}
      \put(60,70){\scriptsize\color{blue}\babar}
      \put(55,65){\scriptsize\color{blue}preliminary}
      \end{overpic}
                            \begin{overpic}[clip,height=2.5in]{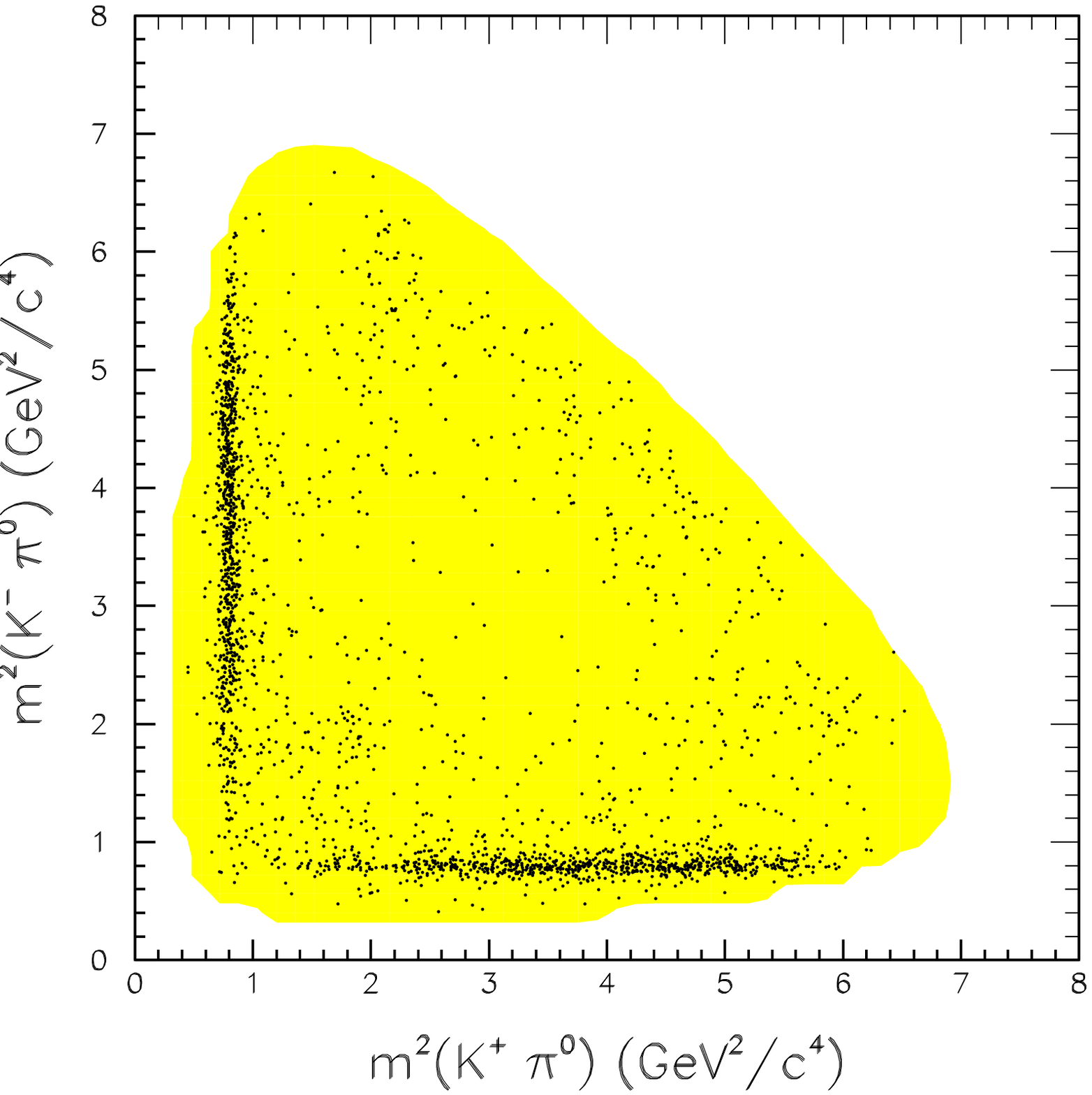}
                        \put(55,75){\scriptsize\color{blue}\babar}
                        \put(50,70){\scriptsize\color{blue}preliminary}
                             \end{overpic}
      \caption{(Left) $\jpsi \ \to \ \pip \pim \piz$   and (Right) $\jpsi \ \to \ \Kp \Km \piz$ Dalitz plots.}
\label{fig:fig6}
 \end{figure}

\begin{figure}
   \centering
                    \begin{overpic}[clip,height=2.5in]{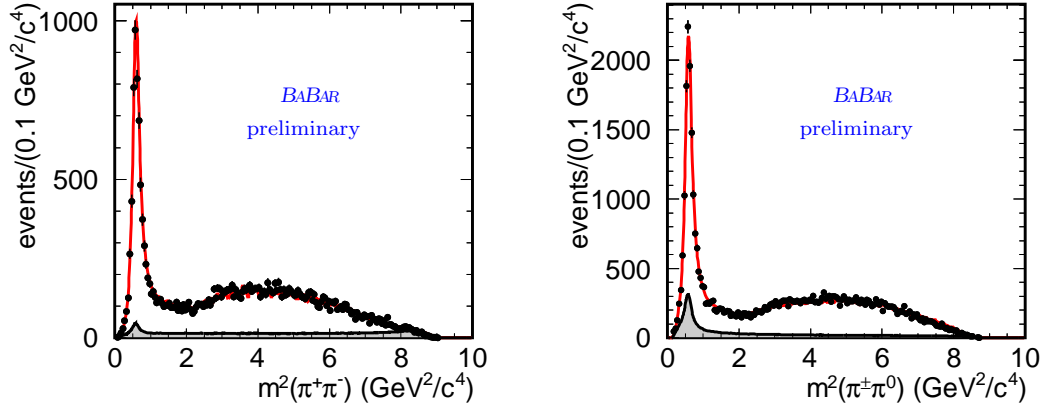}
      \put(25,30){\scriptsize\color{blue}\babar}
      \put(22,27){\scriptsize\color{blue}preliminary}
      \put(75,30){\scriptsize\color{blue}\babar}
      \put(72,27){\scriptsize\color{blue}preliminary}      
      \end{overpic}
            \caption{(Left) $m^2(\pip \pim)$ and (Right) $m^2(\pi^{\pm} \piz)$ for $\jpsi \ \to \ \pip \pim \piz$. Shaded is the background interpolated from $\jpsi$ sidebands.}
\label{fig:fig7}
 \end{figure}

\begin{figure}
  \centering
                      \begin{overpic}[clip,height=2.0in]{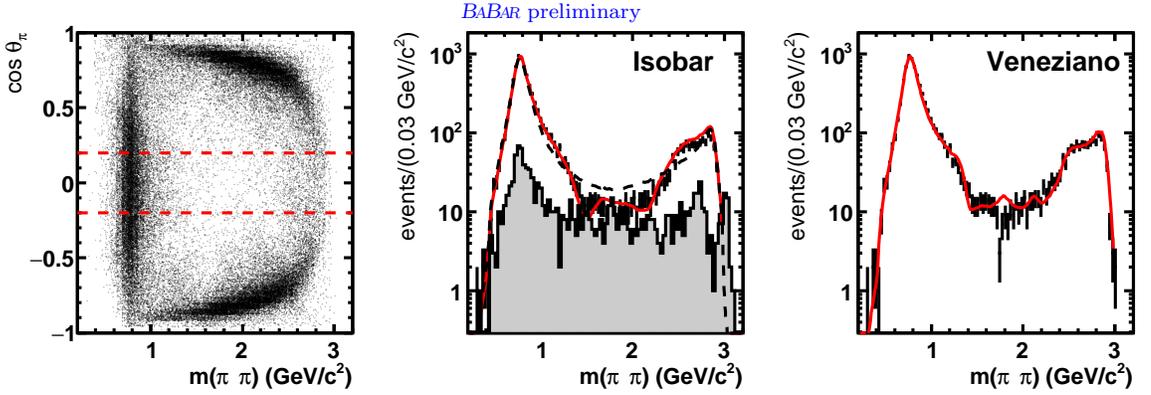}
      \put(40,32){\scriptsize\color{blue}\babar\ preliminary}
      \end{overpic}
      \caption{(Left) Combinatorial $\pi$ helicity angle vs. $m(\pi \pi)$. (Center) and (Right) $m(\pi \pi)$ mass projection for $|cos \theta_{\pi}|<0.2$ for the isobar model fit and Veneziano model in log scale.}
\label{fig:fig8}
\end{figure}

 The Dalitz plot for $\jpsi \ \to \ \pip \pim \piz$ is shown in Fig.~\ref{fig:fig6}(Left) and is dominated by 
 three $\rho(770) \pi$ contributions.
 We perform a Dalitz plot analysis using the isobar model with amplitudes described by Zemach tensors~\cite{zem} and the Veneziano model~\cite{adam}.
    \begin{table}
\begin{center}
\vskip -0.2cm
\begin{tabular}{lr@{}c@{}lr@{}c@{}l|c}
\hline
 \noalign{\vskip2pt}
Final state & \multicolumn{3}{c}{Isobar fraction \%} & \multicolumn{3}{c}{Phase (radians)} & Veneziano fraction \% \cr
\hline
\hline
 \noalign{\vskip2pt}
$\rho(770) \pi$  & 119.0 $\pm$ & \, 1.1 $\pm$ & \, 3.3 & 0. &  & & 120.0 $\pm$  1.9\cr
$\rho(1460) \pi$  & 16.9 $\pm$ & \, 2.0 $\pm$ & \, 3.1 &  3.92 $\pm$ & \, 0.05 $\pm$ & \, 0.11 & 1.53 $\pm$ 0.13\cr
$\rho(1700) \pi$ & 0.1 $\pm$ & \, 0.1 $\pm$ & \, 0.2 &  1.01 $\pm$ & \, 0.35 $\pm$ & \, 0.79 & 0.84 $\pm$ 0.08\cr
$\rho(2150) \pi$ & 0.04 $\pm$ & \, 0.05 $\pm$ & \, 0.02 & 1.89  $\pm$ & \, 0.30 $\pm$ & \, 0.48 & 2.03 $\pm$ 0.17 \cr
$\rho_3(1690) \pi$ & & & & & & & 0.09 $\pm$ 0.02 \cr
 \noalign{\vskip1pt}
\hline
 \noalign{\vskip1pt}
Sum & 136.0 $\pm$ & \,2.3  $\pm$ & \, 4.3 &  & & & 124.5 $\pm$ 2.3 \cr
$\chi^2/\nu$ & & \, 764/552 & \,& & & & 780/554\cr
\hline
\end{tabular}
\caption{Results from the Dalitz analysis using the isobar model (Left) and Veneziano model (Right).}
\label{tab:tab3}
\end{center}
    \end{table} 
  The $\pip \pim$ and $\pi^{\pm} \piz$ squared mass projections are shown in Fig.~\ref{fig:fig7} together with the results from
  the Dalitz analysis.
    The Veneziano model deals with trajectories rather than single resonances.
 The complexity of the model is related to $n$, the number of Regge trajectories included in the fit which requires n=5.
 Fig.~\ref{fig:fig8}(Left) shows the combinatorial $\pi$ helicity angle vs. $m(\pi \pi)$.
   Fig.~\ref{fig:fig8} also shows the $m(\pi \pi)$ mass projection for $|cos \theta_{\pi}|<0.2$ for the isobar model fit (Center) and Veneziano model (Right). The helicity cut removes the $\rho$ reflections enhancing the true $\rho$ signals.
\begin{figure}
  \centering
                          \begin{overpic}[clip,height=2.5in]{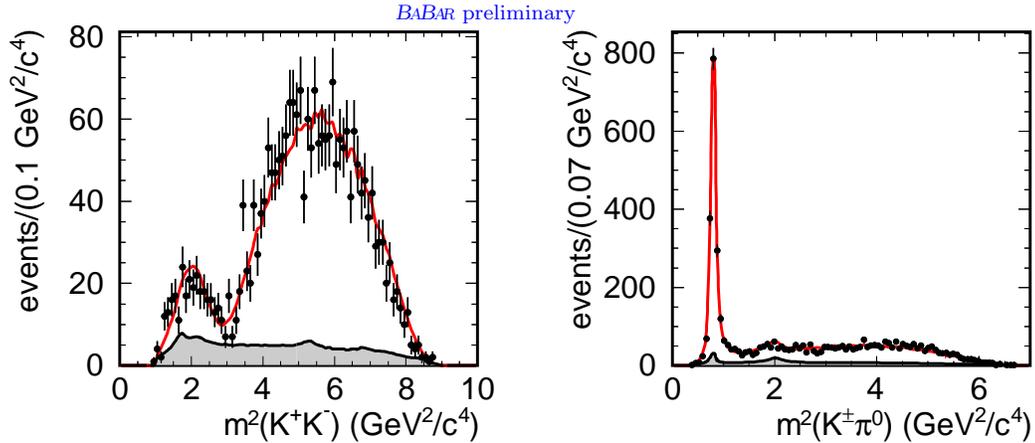}
                        \put(35,40){\scriptsize\color{blue}\babar\ preliminary}
                          \end{overpic}
                                      \caption{Dalitz plot projections with fit results for $\jpsi \ \to \ \Kp \Km \piz$. Shaded is the background interpolated from $\jpsi$ sidebands.}
\label{fig:fig9}
\end{figure}
    \begin{table}
\begin{center}
\begin{tabular}{lcc}
\hline
Final state & fraction \% & phase\cr
\hline
$K^*(892) K$ & 87.8 $\pm$ 2.0 $\pm$ 1.7 & 0. \cr
$\rho(1450)^0\piz$ & 11.5 $\pm$ 2.1 $\pm$ 2.1 & -2.81 $\pm$ 0.25 $\pm$ 0.36\cr
$K^*(1410)K$ & 1.7 $\pm$ 0.7 $\pm$ 1.1 & 2.89 $\pm$ 0.35 $\pm$ 0.08\cr
$K^*_2(1430)K$ & 3.8 $\pm$ 1.4 $\pm$ 0.5 & -2.42 $\pm$ 0.22 $\pm$ 0.07\cr
$\rho(1700)^0\piz$ & 0.9 $\pm$ 1.0 $\pm$ 0.6 & 1.06 $\pm$ 0.20 $\pm$ 0.7\cr
\hline
Total & 105.6 $\pm$ 3.4 $\pm$ 3.0  & \cr
 & $\chi^2/\nu = 94/92$ & \cr
\hline
\end{tabular}
\caption{Results from the $\jpsi   \to \ \Kp \Km \piz$ Dalitz plot analysis.}
\label{tab:tab4}
\end{center}
\end{table}
The fitted fractions and phases from the Dalitz analyses are summarized in Table~\ref{tab:tab3}.
The two models give almost similar data representation, but different fractions.
    
 \section{$\jpsi \ \to \ \Kp \Km \piz$ Dalitz plot analysis}
 The $\jpsi   \to \ \Kp \Km \piz$ Dalitz plot is shown in Fig.~\ref{fig:fig6} (Right) and evidences 
clear $K^{*+}(892)$ and $K^{*-}(892)$ bands with a broad structure in the low $\Kp \Km$ mass region.
The results from the Dalitz analysis using the isobar model are given in Table~\ref{tab:tab4}.

The $\jpsi   \to \ \Kp \Km \piz$  Dalitz projections, with the results from the fit, are shown in Fig.~\ref{fig:fig9}.

    We find the parameters of the low mass $\Kp \Km$ structure in the $\jpsi \ \to \ \Kp \Km \piz$ are consistent for being associated to $\rho(1450)$ also observed in $\jpsi \ \to \ \pip \pim \piz$. 
 Combining the two measurements we obtain
   \begin{center}
    $\frac{\BR(\rho(1450)^0 \ \to \ \Kp \Km)}{\BR(\rho(1450)^0 \ \to \ \pip \pim)} = 0.190 \pm 0.042 \pm 0.049$.
      \end{center}

%%%%%%%%%%%%%%%%%%%%%%%%%%%%%%%%%%

\end{document}

%% file: econfmacros.tex
\def\beq{\begin{equation}}
\def\eeq#1{\label{#1}\end{equation}}
\def\eeqn{\end{equation}}

\def\beqa{\begin{eqnarray}}
\def\eeqa#1{\label{#1}\end{eqnarray}}
\def\eeqan{\end{eqnarray}}

\let\bar=\overbar

\def\etal{{\it et al.}}

\def\Dslash{\not{\hbox{\kern-4pt $D$}}}
\def\dslash{\not{\hbox{\kern-2pt $\del$}}}

\def\BR{\mbox{\rm BR}}

\def\msb{{\bar{\ssstyle M \kern -1pt S}}}

%% file: charm2015_APalanoBabar.bbl
\begin{thebibliography}{99}
\bibitem{conj}
Charge conjugation is implied through all this work.
\end{thebibliography}
